# Electric-Dipole Effect of Defects on Energy Band Alignment of Rutile and Anatase $TiO_2$

Daoyu Zhang,[1] Minnan Yang,[2] Shuai Dong[1,*]

[1]*Department of Physics & Jiangsu Key Laboratory for Advanced Metallic Materials,, Southeast University, Nanjing, 211189, China*

[2]*Department of Physics, China Pharmaceutical University, Nanjing, 211198, China*

**ABSTRACT:** Titanium dioxide materials have been studied intensively and extensively due to photocatalytic applications. A long-standing open question is the energy band alignment of rutile and anatase $TiO_2$ phases, which can affect the photocatalytic process in the composite system. There are basically two contradictory viewpoints about the alignment of these two $TiO_2$ phases supported by respective experiments: 1) straddling type and 2) staggered type. In this work, our DFT plus *U* calculations find that the perfect rutile (110) and anatase (101) surfaces have the straddling type band alignment, whereas the surfaces with defects can turn the band alignment into the staggered type. The electric dipoles induced by defects are responsible for the reversal of band alignment. Thus the defects introduced during preparations and post-treatment processes of materials are probably the answer to above open question regarding the band alignment, which can be considered in real practice to tune the photocatalytic activity of materials.

[*] Email: sdong@seu.edu.cn

## I. INTRODUCTION

The discovery of the water photolysis on the TiO$_2$ electrode by Fujishima and Honda [1] have evoked enormous amount of investigations on TiO$_2$ [2]. During the past four decades, a wealth of information related with photocatalytic properties of TiO$_2$, as well as other physical and chemical properties, has been collected [3, 4]. Rutile and anatase are the two principal crystalline phases of TiO$_2$ quite suitable for the photocatalytic applications. It is widely assumed that the anatase phase TiO$_2$ displays higher photocatalytic activity than the rutile one, because anatase materials have lower rates of recombination of electron-hole pairs.

Most interestingly, the composite consisting of anatase and rutile TiO$_2$ exhibit even higher photocatalytic activity than individual components due to the synergistic effect on the separation of excited electrons and holes at the interface between the anatase and rutile phases [5-10]. A lot of previous experimental works were devoted to probe the migration direction of carriers at the interface. However, two opposite results have been obtained: 1) electrons transfer from anatase to rutile [11-13] and 2) electrons transfer from rutile to anatase [14-17].

The debate on this charge migration also took place in the theoretical aspect. The effective separation of the photoexcited electron-hole pairs at the interface is supposed to be the result of the energy difference of band edges of anatase and rutile. Two types of the band alignment of anatase and rutile phases were predicted by using different theoretical methods, leading to two opposite directions of electron transfer, as illustrated in Fig. 1. The straddling type (Fig. 1(a)) is characterized by band edges of anatase straddling those of rutile, which will drive the migrations of both electrons and holes from anatase to rutile [18]. For the staggered type (Fig. 1(b)), the band edges of anatase are lower than those of rutile, leading to the inverse electrons/holes migrations [17, 19-21].

To date, the scenario of carrier transfer process in the mixed-phase TiO$_2$ composite remains ambiguous, which seriously influences the correct design of the mixed-phase TiO$_2$ to improve the photocatalytic activity of this material. Therefore, it is physical interesting and application meaningful to figure out the real mechanisms, even partial, involved in the band alignment of rutile and anatase TiO$_2$.

At the interface between two semiconductors, many factors, such as the charge transfer across the interface, dangling bonds, atomic arrangements at the interface, charge trapping sites, the interfacial tense, the interfacial orbital reconstruction, influence the energy band alignment of the heterostructure,[20, 22-24] so it is difficult to extract the wanted information of the effect of the electric dipole just induced by the interfacial defects on the band alignment. Thus, the model of the interface between two TiO$_2$ phases is not suitable to act as the calculational method for approaching the aim of this article research.

To clearly understand the electric dipole effect of defects on the band alignment of the rutile and anatase TiO$_2$, we carry out computational analyses separately on the two phases, obtaining their absolute band energies and band alignment. In this way, our provided difference between conduction band edges of two TiO$_2$ phases, namely the band offsets, is the Schottky limit value. The Schottky limit is an important parameter that acts as the boundary conditions imposed on a particular interface, and one can provides just corrections to the Schottky limit to get the band offsets of the real heterostructure. In this sense, in the following, the discussion on the transfer of the photoexcited carriers is based on the Schottky limit.

Our previous works [25, 26] and works by other groups [27, 28] predicted that the electric dipoles created by chemisorbed molecules or atoms on the surface of a semiconductor can significantly change the band-edges energies. Based on this idea, in the present work, the bridging oxygen vacancies (O-vac's) and the hydroxyl groups (O-H's), which can be introduced into the TiO$_2$ surfaces during the material preparation, are studied to verify the effect of electric dipoles on the band alignment of the rutile and anatase TiO$_2$.

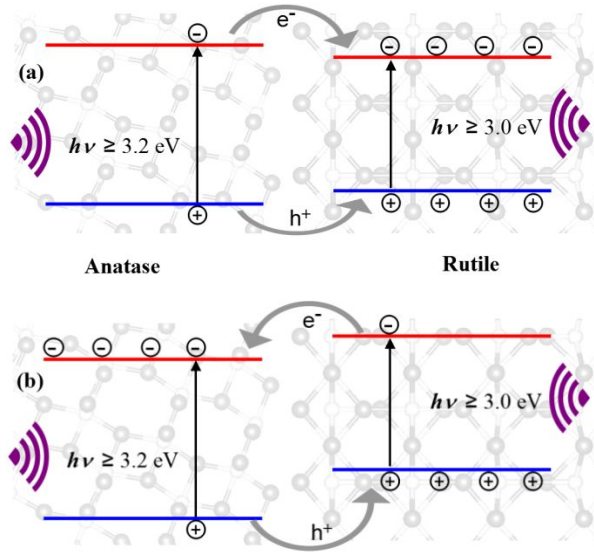

**Fig. 1** The proposed two types of band alignment between rutile and anatase TiO$_2$: (a) the straddling type, in which excited electrons and holes will prefer to accumulate in the rutile phase; and (b) the staggered type, in which the excited electrons prefer to migrate to the conduction band of the anatase while the holes prefer to move to the valence band of the rutile.

## II. MODEL & METHOD

The first-principles calculations were performed using the projector-augmented wave (PAW) pseudopotentials as implemented in the Vienna *ab initio* Simulation Package (VASP) [29, 30]. The Perdew-Burke-Ernzerhof (PBE) GGA exchange-correlation functional was used. The Hubbard-type correction ($U$) within Dudarev's approximation [31] was applied to strongly localized Ti's 3$d$ orbitals for remedying on site Coulomb interaction. The energy cutoff for plane wave basis was set to be 450 eV and the convergence criteria in energy were $10^{-5}$ eV. The atomic positions were relaxed towards equilibrium using the conjugate gradient method until the force on each atom is less than 0.01 eV/Å. Gaussian smearing with a width of 0.01 eV was employed for calculating partial occupancies.

The stoichiometric $p(3\times2)$ rutile TiO$_2$ (110) and $p(1\times3)$ anatase TiO$_2$ (101) supercell surfaces were built from experimental lattice parameters. The two supercells have the same number of 144 atoms, half of which are fixed at their bulk positions during the relaxation process, as indicated in Fig. 2(a) and 2(b). The numbers of two-fold bridging O atoms and five-fold bridging Ti atoms are also same at these two surfaces, beneficial to compare the following calculated results of the two surfaces. In real materials, these two surfaces are also the most stable and common ones [32]. A vacuum space of about 11 Å was set for separation of the surface slab from its periodic images. In the direction perpendicular to the slab, the monopole, dipole, and quadrupole corrections have been applied to the electrostatic interaction between the slab and its periodic images. $\Gamma$-point-only sampling was used for the geometrical relaxation of surfaces. Automatically generated $\Gamma$-point-centered 3×2×1 (rutile) and 2×2×1 (anatase) Monkhorst-Pack meshes were used for static electronic structure calculations.

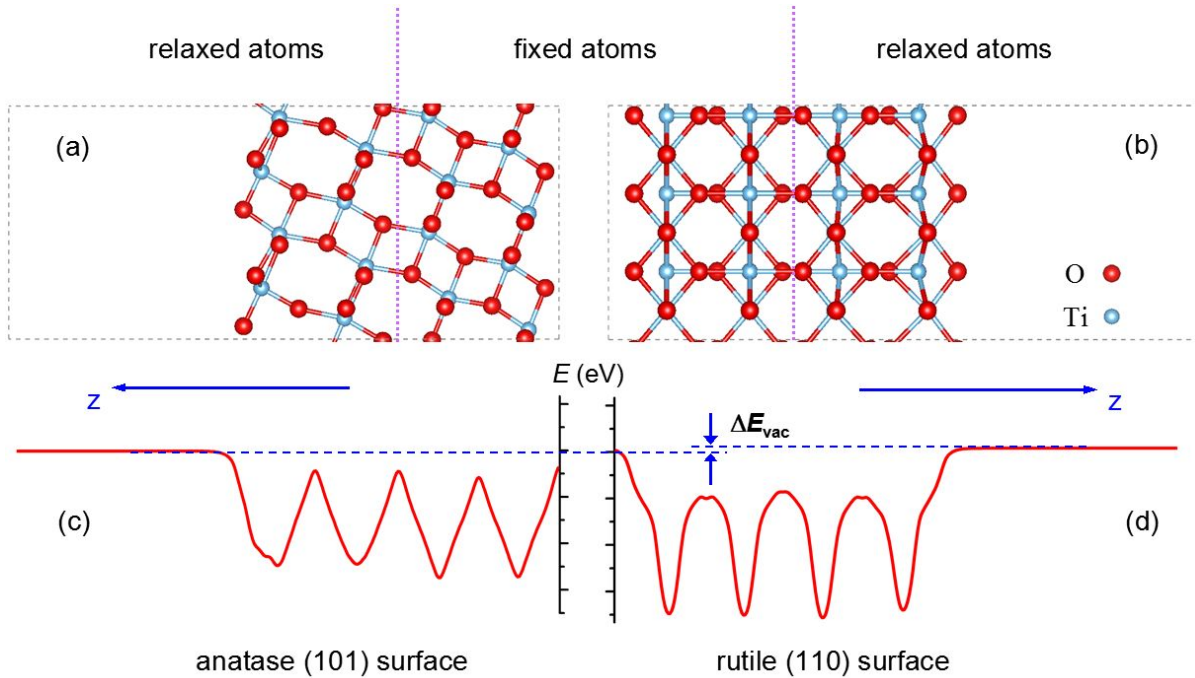

**Fig. 2** The surface models of (a) anatase (101) and (b) rutile (110), both of which possess the same numbers of two-fold bridging O atoms and five-fold bridging Ti atoms. The sketch of (x,y)-planar averaged electrostatic potential for a pair of (c) anatase (101) and (d) rutile (110) surfaces.

According to previous literature, there are several approaches to align band energy such as the vacuum level alignment, the charge neutrality level alignment, the common anion rule, and so on [33, 34]. Although the band alignment deduced from interface supercell model can reveal accurate values for the band offsets [35], here the surface model using the vacuum level as a common energy reference is selected. This choice can obtain the relative values between band edges of rutile and anatase and effectively reduce inaccuracy of the offsets [36]. In details, the strategy for the band alignment of corresponding anatase and rutile surfaces is as following. First, based on the (x,y)-planar average electrostatic potential [21], the difference of the deep vacuum space between the pair of surfaces is calculated as $\Delta E_{vac} = E_{vac}(rutile) - E_{vac}(anatase)$. Second, the Kohn-Sham valence band edges from the DFT+$U$ calculations are aligned by subtracting $\Delta E_{vac}$. Third, the conduction band edges is aligned based on the above aligned valence band edges by adding the commonly accepted band gaps of anatase (3.2 eV) and rutile (3.0 eV) $TiO_2$ [37].

**III. RESULTS & DISCUSSION**

First, we focus on the clean surfaces. Our DFT+$U$ calculations show that the stoichiometric rutile (110) surface is almost nonpolar with the very tiny dipole moment of only ~0.03 eÅ, while the stoichiometric anatase (101) is highly polar with a distinct moment of ~0.29 eÅ independent of the value of $U$ (as listed in the second column of Fig. 3). This electric dipole occurring at the anatase (101) surface is helpful to separate the photogenerated electron-hole pairs and block the electron-hole recombination. The direction of the electric dipole moment from the surface to the inner implies that photogenerated holes will gather at the surface while electrons will migrate to inner side. This intrinsic electric dipole at the polar anatase (101) surface may be the reason for mitigating rates of recombination and back-reaction comparing with the rutile phase $TiO_2$ [38-40]. As a consequence, anatase phase $TiO_2$ displays better photocatalytic activity.

The second column of Fig. 3 shows the energy band alignment for clean surfaces of anatase and rutile phases. The conduction and valence band edges of the anatase (101) surface straddle those of the rutile (110) surface, in agreement with the calculated result of a quantum-dot supercell composing of anatase and rutile pair [18]. In the case of the straddling type, excess electrons and holes made by radiation will accumulate in the conduction band and valence band of rutile $TiO_2$ respectively, provided anatase and rutile keep in close contact with each other. Because rutile $TiO_2$ exhibits high rates of recombination [41], the accumulated electrons and holes may quickly recombine with each other before they move to the reactants adsorbed on the surfaces, thus the photocatalytic activity of mixed anatase and rutile phases would be expected be low efficient with the straddling type alignment. In this sense, the mixed-phase $TiO_2$ materials free of defects are not advantageous for photocatalytic applications.

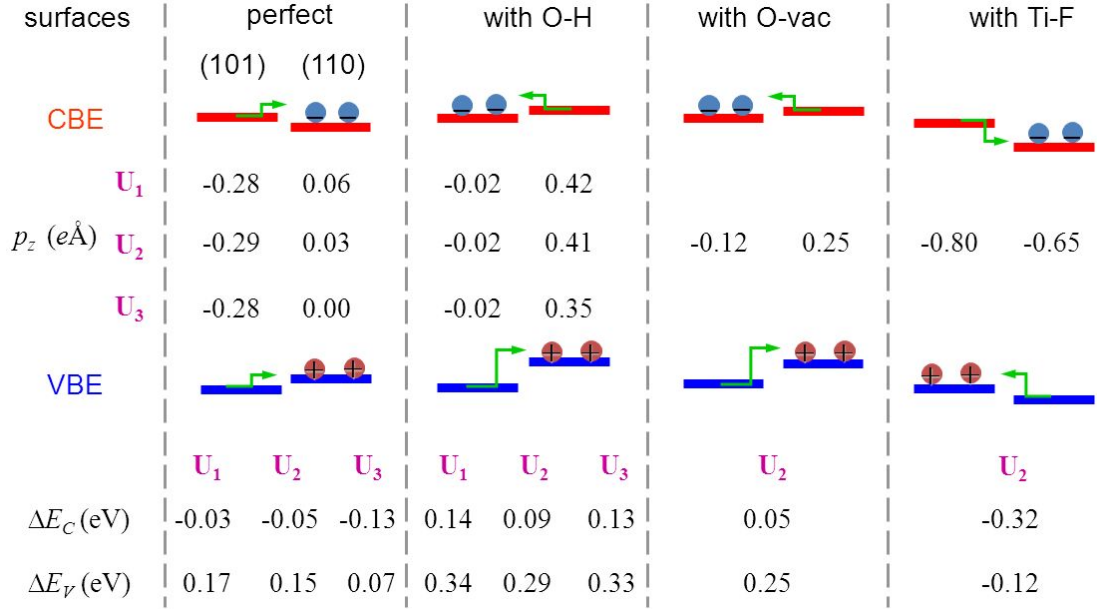

**Fig. 3** The DFT+$U$ calculated relative band edges of the clean surfaces and surfaces with defects of O-H, O-vac, and Ti-F. CBE (VBE) denotes the conduction (valence) band edge, and $\Delta E_C$ ($\Delta E_V$) is the difference of the conduction (valence) band edges between the rutile (110) and anatase (101) surfaces. $p_z$ denotes the z component (perpendicular to the surface) of the electric dipole moment of the surface. The positive value of $p_z$ means that its direction is pointing away from the surface. Here $U_1$ = 3.3 eV; $U_2$ = 4.3 eV; $U_3$ = 5.3 eV.

In fact, TiO$_2$ materials always have defects, depending on the preparation conditions and the post-treatment processes. When sputtered and annealed in ultra-high vacuum or bombarded with electrons, TiO$_2$ samples will lose some bridging oxygen atoms forming oxygen vacancies (O-vac's) [32]. When treated in the hydrogen containing atmosphere, they are ready to combine with hydrogen atoms forming hydroxyl groups (O-H's) [42, 43]. The energy band alignment between the anatase (101) and rutile (110) surfaces with the O-H and O-vac coverage of 1/6 monolayer is shown in the third and fourth columns of Fig. 3, respectively. Such a pair of surfaces with equivalent defects has the band alignment of the staggered type, independent of the value of $U$. The switch of the band alignment type from the straddling type to the staggered one indicates that the effect of defects play a crucial role to tune the band alignment and thus the photocatalytic activity of two TiO$_2$ phases and their composition.

The effect of the defects on the band alignment is related with the electric dipoles introduced by defects themselves. Previous works had demonstrated that the chemisorbed functional groups on semiconductors can supply excess electric dipoles, which change electron energies in semiconductors and shift their whole energy bands together [26, 28]. And the energetic variation of electrons in a semiconductor $\Delta E_{\text{dip}}$ can be formulated within the parallel-plate capacitor approximation as:

$$\Delta E_{\text{dip}} = e \frac{\Delta p_z}{A \varepsilon \varepsilon_0} \quad (1)$$

where $A$ is the surface area, $\varepsilon$ is the effective dielectric constant of the surface layer, and $\Delta p_z$ is the electric dipole moment induced by functional groups. $\varepsilon_0$ is the dielectric constant of vacuum and $e$ is the elementary charge. Electron energies vary linearly with $\Delta p_z$. The defects of O-H's (O-vac's) bring forth the dipole moments of 0.38 $e$Å (0.22 $e$Å) and 0.27 $e$Å (0.17 $e$Å) for the rutile (110) and anatase (101) surfaces respectively. In other words, the rutile (110) surface with defects has a larger $\Delta p_z$ than the anatase (101) surface with the "identical" defects, thus $\Delta p_z$ increases electron energies in the former more than that in the latter. Even more possibly, the conduction band edge of the rutile (110) surface would surpass that of the anatase (101) surface, such as our cases studied here.

Why do the rutile (110) and anatase (101) TiO$_2$ phases with the same defect coverage possess different $\Delta p_z$'s? The case with the same hydroxyl coverage (1/6 monolayer) will be taken as an example to reveal the underlying physical mechanism. First, the Bader charges [44] on H and O atoms of the hydroxyl groups at the anatase (101) and rutile (110) surfaces are almost identical. The difference of charges on H (O) atoms at two surfaces is only 0.0001 (0.0351) $e$, whose contribution to $\Delta p_z$ is negligible. Second, the bond lengths of O-H's at the two surfaces are identical, being 0.968 Å. Third, according to previous studies, the intrinsic electric dipole of the adsorbed hydroxyl groups themselves plays a dominant role in shifts of the band edges of TiO$_2$ with respect to the polaronic dipole created by structural distortion and charge rearrangement [25]. Thus, one can infer that the different configurations of the hydroxyl groups at the two surfaces are responsible for their different $\Delta p_z$. As shown in Fig. 4, the hydroxyl group at the relaxed rutile (110) surface is almost vertical, i.e. the angle between the hydroxyl and the normal direction of surface is only 1.1°. Whereas, the hydroxyl group at the relaxed anatase (101) surface is tilted with the angle of 25.7°. Taking the dipole moment of O-H ($p_{\text{O-H}} = 0.32$ $e$Å) estimated from the dipole moment of a water molecule [25], the projection of $p_{\text{O-H}}$ to the normal direction is 0.28 and 0.32 $e$Å for the anatase (101) and rutile (110) surfaces respectively, in good agreement with $\Delta p_z$ (0.27 and 0.38 $e$Å) obtained in our DFT+$U$ calculation. This agreement supports that the dipole moment of polar groups adsorbed on surfaces is an important source to tune electron energies in TiO$_2$ and the energy band alignment between the different TiO$_2$ phases.

To further examine the effect of higher defect coverage on the band alignment type, the rutile (110) and anatase (101) TiO$_2$ surface with the 1/3, 2/3, 1 monolayer hydroxyl coverages have also been calculated. The obtained dipole moment is in linear proportional to the hydroxyl coverage, which is always larger for the rutile surface than the anatase one at the same coverage. Considering Eq. 1, the band edges of the rutile (110) surface always surpass those of the anatase

(101) surface upon the same extent of hyroxylation from 1/6 ML to 1 ML, maintaining the staggered type alignment.

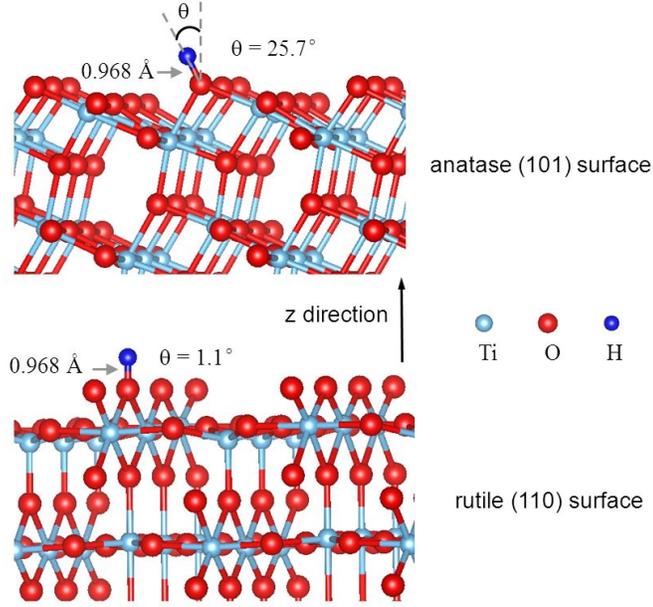

**Fig. 4** The configurations of the hydroxyl groups at the anatase (101) and rutile (110) surfaces.

The O-H's and O-vac's introduce positive $\Delta p_z$'s into TiO$_2$ surfaces. Then an interesting question is what will happen if a negative $\Delta p_z$ is introduced. This supposition can be tested through adsorption of fluorine on the rutile (110) and anatase (101) TiO$_2$ surfaces. The fluorine atom has a great electronegativity value and readily forms polar covalent bond with an under-coordination Ti atom [45, 46]. Our calculation finds that in this case electron energies in the rutile TiO$_2$ decrease more than that in the anatase TiO$_2$, then the reverse straggled type (*i.e.* band edges of anatase are higher than those of rutile) may take place, as illustrated in the last column of Fig. 3. Negative $\Delta p_z$'s are introduced with the values of -0.51 $e$Å and -0.68 $e$Å for the anatase (101) and rutile (110) surfaces respectively when $U$ = 4.3.

**Table 1** The difference in the total energy between the surfaces with and without the defects in unit of eV, $\Delta E$(defect), calculated using the DFT+$U$ method.

|  | $U$ = 3.3 eV | | $U$ = 4.3 eV | | $U$ = 5.3 eV | |
|---|---|---|---|---|---|---|
| Surface | (101) | (110) | (101) | (110) | (101) | (110) |
| $\Delta E$(O-H) | -3.40 | -3.69 | -3.60 | -3.86 | -3.95 | -4.16 |
| $\Delta E$(O-vac) | 9.28 | 8.99 | 9.01 | 8.77 | 8.51 | 8.34 |
| $\Delta E$(Ti-F) |  |  | -2.89 | -3.51 |  |  |

Above calculations have shown that the same defect coverage (1/6 monolayer of O-H's or O-vac's) on the anatase (101) and rutile (110) surfaces can lead the band alignment to the staggered type. However, in practice, the two TiO$_2$ surfaces may own different defect coverage under the same preparation conditions. According our DFT+$U$ calculationt, the rutile (110) surface favors the defects relative to the anatase (101) surface. As shown in Table 1, the energy increments from defects are lower for the rutile (110) surface than for the anatase (101) one, which will lead to different defect coverage in real materials.

In this sense, higher defect coverage on the rutile (110) surface can magnify the difference of $\Delta p_z$ between two phases of TiO$_2$, and further enhance the staggered type energy band alignment. Taking an extreme case for an example, for the rutile (110) surface covered with 1/6 monolayer O-H and the clean anatase (101) surface, the valence (conduction) band edge of the rutile is higher for 0.73 (0.54) eV than that of anatase. Such significant energy spaces between corresponding band edges for two TiO$_2$ phases also was observed by experiments. The polycrystalline anatase thin films and rutile single crystals prepared by Pfeifer *et al.* shows that the VBM (CBM) of rutile is 0.7 (0.5) eV above that of anatase according to the photoelectron spectroscopy analysis [47].

It should be noted that the type of energy band alignment between the rutile and anatase TiO$_2$ is also dependent on the measuring methods. Noting that the (photo)electrochemical techniques derives the band edges from the flatband potential, where the band bending at the semiconductor-liquid interface is eliminated. So, unlike the photoelectron spectroscopy method, which measures values with the energy-bands shift arisen from the surface dipoles induced by defects, those electrochemical methods should give the same band edges of a semiconductor despite of the surficial band bending caused by defects. This is true as seen by the following evaluation. According to our DFT+U (U = 4.3 eV) calculations the mean value of energy-bands shift per unit of the dipole moment for anatase (rutile) is 1.57 (1.65) eV/$e$Å. And combined with data in Figure 3, when the band bending caused by defects becomes completely flat, the conduction band edge of anatase is higher than that of rutile of 0.56 eV and 0.62 eV for surfaces with zero and 1/6 monolayer hydroxyl coverage respectively. In fact, in the electrochemical experiment conducted by Kavan *et al* [37], electrodes prepared from anatase crystals had the (101) face exposed, and were annealed in hydrogen atmosphere at 500-600 °C to adsorb lots of hydroxyl groups [48], and their impedance analysis established that the flatband potential of the anatase (101) surface is ~0.2 eV higher than that of the rutile electrode prepared under the same conditions, lower than our estimate value of 0.62 eV from ideal plat band potential. Thus the electrochemical methods did not give a staggered type of the band alignment for rutile and anatase TiO$_2$ in the presence of defects at surfaces, rather a straddling type.

## IV. CONCLUSION

The DFT+$U$ calculations have shown that the energy band alignment for the perfect anatase (101) and rutile (110) surfaces is the straddling type, whereas the two surfaces with defects have the staggered band alignment. The common reductant defects O-H's and O-vac's, as well as oxidative Ti-F's, prefer the staggered type band alignment. The switch of the band alignment from the straddling to the staggered is attributed to the electric dipoles induced by defects. Our computational results can provide a reasonable explanation to the long-standing debate on the energy band alignment for rutile and anatase $TiO_2$ and shed light to the electric-dipole effect tuning of the photocatalytic activity.

## ACKNOWLEDGMENT

Work was supported by the National Natural Science Foundation of China (Grant Nos. 51322206 and 11274060) and the Jiangsu Key Laboratory for Advanced Metallic Materials (Grant No. BM2007204).